\documentclass[a4paper,10pt]{article}

\usepackage[utf8]{inputenc}

\usepackage{cite}
\usepackage{amsmath,amssymb,amsfonts}
\usepackage{algorithmic}
\usepackage{graphicx}
\usepackage{textcomp}
\usepackage{xcolor}

\usepackage{url}
\usepackage{multirow}
\usepackage{booktabs}
\usepackage{subcaption}

\usepackage[margin=1.5in]{geometry}

\def\BibTeX{{\rm B\kern-.05em{\sc i\kern-.025em b}\kern-.08em
    T\kern-.1667em\lower.7ex\hbox{E}\kern-.125emX}}

\begin{document}

\title{Studying the workload of a fully decentralized Web3 system: IPFS}


\author{Pedro Ákos Costa \\ NOVALINCS \& NOVA University of Lisbon \\ pah.costa@campus.fct.unl.pt \and

João Leitão \\ NOVALINCS \& NOVA University of Lisbon \\ jc.leitao@fct.unl.pt \and

Yannis Psaras \\ Protocol Labs \\ yiannis@protocol.ai}


\maketitle

\begin{abstract}

Web3 aims at creating a decentralized platform that is competitive with modern
cloud infrastructures that support today's Internet. However, Web3 is still
limited, supporting only applications in the domains of content creation and
sharing, decentralized financing, and decentralized communication. This is
mainly due to the technologies supporting Web3: blockchain, IPFS, and libp2p,
that although provide a good collection of tools to develop Web3 applications,
are still limited in terms of design and performance. This motivates the need to
better understand these technologies as to enable novel optimizations that can
push Web3 to its full potential. Unfortunately, understanding the current behavior of
a fully decentralized large-scale distributed system is a difficult task, as there is
no centralized authority that has full knowledge of the system operation.

To this end, in this paper we characterize the workload of IPFS, a key enabler
of Web3. To achieve this, we have collected traces from accesses performed by
users to one of the most popular IPFS gateways located in North America for a
period of two weeks. Through the fine analysis of these traces, we gathered the
amount of requests to the system, and found the providers of the requested
content. With this data, we characterize both the popularity of requested and
provided content, as well as their geo-location (by matching IP address with the
MaxMind database). Our results show that most of the requests in IPFS are only
to a few different content, that is provided by large portion of peers in the
system. Furthermore, our analysis also shows that most requests are provided by
the two largest portions of providers in the system, located in North America
and Europe. With these insights, we conclude that the current IPFS architecture
is sub-optimal and propose a research agenda for the future.

\end{abstract}

%
%


\section{Introduction} 

Today's Internet is supported by large cloud providers that include Google,
Amazon, and Microsoft. These providers support a wide variety of web services
that operate at a large scale serving huge numbers of users scattered
throughout the World. Nevertheless, this paradigm forces application providers
to fully rely and trust on centralized cloud infrastructure that are susceptible
to attacks\,\cite{cloudattack1,cloudattack2} and dictate the terms of service
with little to no competition. Moreover, in most application scenarios the
control of user data is relinquished in some way to these operators, which is
undesirable (and being targeted by legislation such as the European GDPR). To
address this, and partially achieved by the increased popularity and use cases
enabled by blockchain technologies\,\cite{eth,bitcoin}, the concept of
Web3\,\cite{web3} emerged. Web3 aims at decentralizing web technologies to
improve user security and privacy as well as providing guaranteed ownership of
content, through the use of existing novel peer-to-peer
protocols\,\cite{p2p,leitao:msc}.

However, Web3 is still in its early stages and is yet to become competitive with
modern cloud infrastructures, in terms of flexibility, application development,
security and privacy, and performance. This is due to the current Web3 main
technology enablers: blockchain, that maintains and replicates the system state;
libp2p\,\cite{libp2p}, that is used to develop decentralized applications
(dApps) and their support; and IPFS\,\cite{_ipfs,ipfs}, that is used as an entry-point
to most dApps, that are still restricted to the domains of content creation and
sharing\,\cite{fleek,nftstorage}, decentralized financing\,\cite{uniswap},
decentralized communication\,\cite{berty,discussify}, among a few others. This
is because, blockchain, although important for
decentralization, also limits the amount of interactions applications can
have, as these are mostly made through the blockchain. Furthermore, IPFS still
has a large space for performance improvements, as recent studies show that
searching for content in IPFS can take up to $2,5$ hours\,\cite{consenSysMesurments}.

Most of these technologies rely on peer-to-peer protocols that have been the
subject of a large body of research in the past. However, most Web3 technology
either are unaware of this research (such is the case for most blockchains that resort
to flood dissemination\,\cite{blockchainflood} instead of a more efficient
gossip dissemination\,\cite{plumtree} to disseminate mined blocks) or leverage
it using similar configurations as described in the research, and sometimes
experience performance issues (such is the case of the DHT in IPFS). However, it a
non-trivial task to understand the causes for performance issues and how to
mitigate them on a decentralized large-scale distribute system, as there is no
single node that holds the entire view over the state of the system. Hence, it
is paramount to devise techniques and to analyze data obtained from such systems
to understand the current environment, workload, and system distribution, to then
guide research and optimization for Web3 support\,\cite{pkad}.

In this paper we present an in-depth analysis of the workload on IPFS. We have
gathered two weeks worth of logs of one of the most popular IPFS gateways
located in North America. These logs contained the (access) requests made from
IPFS users across the World to large amounts of content stored in IPFS. We
analyzed these logs and performed the same (valid) requests that were observed
in the logs to fetch the information regarding the providers of the requested
content. This allowed two things: first to correlate which and how many
providers provided which and how many requested content. Second, and through the
use of the MaxMind\,\cite{maxmind} database that matches IP addresses to
geo-location information, to find the relation of the location of the origin of
requests and location of providers of content to those requests. With this data, we
analyzed the IPFS workload and found that most content is provided only by a few
providers that are located on North America and Europe.

The remainder of the paper is structured as follows:
In Section~\ref{sec:ipfs} we provide a brief description of the IPFS network and how it
operates.
In Section~\ref{sec:meth} we detail our methodology to gather and analyze the data.
In Section~\ref{sec:res} we describe our results, providing insights on the workload
of the IPFS network.
In Section~\ref{sec:related} we discuss related work.
Finally, Section~\ref{sec:con} concludes the paper with final remarks and directions for future work.

\section{IPFS}\label{sec:ipfs} 

IPFS is large scale peer-to-peer distributed system that aims at connecting
computing devices through a shared file system. To enable this, IPFS relies on
libp2p\,\cite{libp2p} to handle networking aspects. To connect peers in the
network, libp2p relies on a distributed hash table (DHT), implemented as a variant of
the Kademlia protocol\,\cite{kad}. IPFS leverages this DHT to distribute and
search content and peers in the system. Content in IPFS is immutable, having
each content associated an identifier (\emph{cId}) that is a
multi-hash\,\footnote{A multi-hash encodes not only the hash of the data, but
also hashes of metadata about the hashing process and data type.} of the
content. Similarly, each peer in IPFS also has an identifier (\emph{peerId})
that is a multi-hash of the peer's public key. Peers organize themselves in the
DHT according to a SHA-256 hash of their peerId and store content pointers
according to the SHA-256 hash of the cId. Furthermore, each peer has associated
to it a list of multiaddresses, that describe the internet addresses of the peer
(this can be ipv4, ipv6, dns, or other kind of address).

Note that peers do not store the content itself but only a pointer to the peer
providing the content. As such, for a peer to publish content on IPFS, the
operation the peer effectively performs is to announce that it provides the
content by publishing on IPFS a \emph{provider record}. A provider record
contains a mapping of a cId to peerIds -- its providers. As per the Kademlia
operation, this provider record will be stored in the $k$ closest peers to the
hash of the cId on the DHT. In IPFS $k$ has a value of $20$. Note that the same
content can be provided by multiple peers.

To fetch content IPFS uses a protocol named \emph{Bitswap}\,\cite{bitswap} that
performs both discovery and data transfer. For Bitswap to discover content it
begins to perform a local one-hop flood request for the cId. This will send a message to
all neighboring peers asking if they have the cId locally. If the answer is
positive, Bitswap begins transferring the content with a process akin to
BitTorrent\,\cite{bittorrent}. If the response is negative and the content cannot be found on a
neighboring peer, it resorts to the DHT to find
the provider records of the cId. Once Bitswap has obtain one provider record, it
will start to try to transfer the content from providers on that record.

IPFS has two modes of operating: as a server or as a client node. Server nodes
are (usually) publicly reachable in the Internet and kept in the DHT to enable
routing among peers and serve content to the network. Client nodes connect to
the DHT, but do not maintain the DHT, meaning that client nodes can only perform
requests to the DHT and do not help in the routing process. Additionally, an
IPFS peer can also act as a gateway, in which the IPFS node also runs a web
server that grants access to IPFS via a browser interface to user. In more
detail, a gateway node is able to transform an HTTP request into a valid IPFS
request, that will either trigger a Bitswap and/or DHT operation.

Furthermore, not all IPFS nodes are publicly reachable. This is the case for
nodes that are behind a NAT. In this case, an IPFS node can request a publicly
reachable IPFS node to relay traffic for that node.

\section{Methodology}\label{sec:meth} 

In this section we provide a detailed description of our methodology to study
and characterize the IPFS workload. In summary, we collected two weeks worth
(from March 7th to March 21th of 2022) of logs from one of the most popular
IPFS gateway -- \emph{ipfs.io} -- that is located in North America. These logs
were produced by an nginx HTTP server that logged every HTTP
request made. As such, each entry in the log contains an HTTP request made by a
user to the IPFS gateway.

To process these logs, we filtered all non-valid HTTP requests (e.g.,
\textsc{POST} operations, out of format entries), and extracted the cId in each
valid HTTP request. From this, we fetch all the available provider records
from IPFS for each cId. To obtain geo-location information from both the
requests and providers, we matched the IP addresses we found against the MaxMind
GeoLite2 Free Geolocation Database\,\cite{maxmind}.

Finally, we matched both datasets on the requested and provided cId to produce a
global view that shows where content is requested and where that content is
served. In the following, we describe this process in more detail.

\subsection{Processing the requests}

The first step in analyzing the IPFS workload is to extract the requested cIds.
To do this, we parse the gateway logs into a structured data format that can be
easily processed. The gateway logs are generated by an nginx server, that
effectively acts as a reverse proxy for an IPFS server node that acts as an IPFS
gateway. Each log entry contains information about the received HTTP request by
the IPFS gateway. In particular, we are interested in the following information
of the HTTP requests: the IP address of the requester, to extract geo-location
information; the HTTP operation and the HTTP status, to filter unwanted HTTP
requests for our study (e.g., \textsc{POST} operations and $400$ \textsc{GET}
operation status); and the HTTP target and the HTTP host, that effectively
contain the requested cId.

We begin by filtering logs that are out of format. This amounts to $0.001\%$ of
all requests and include requests that have unparsable HTTP targets which contained
hexadecimal codes in this field. Next, we remove all requests that are not \textsc{GET}
HTTP requests, as these are the only ones that actually make requests to
IPFS. This filters about $19\%$ of all requests contained in the logs, leaving
almost $81\%$ of log entries with only \textsc{GET} operations.

However, not all \textsc{GET} operations are of interest to process the
workload. Such is the case for \textsc{GET} operations that did not succeed
(i.e., where the reply had a HTTP status code of $400$) or that do not contain a
cId in their request. We filter \textsc{GET} operations that did not succeed due
to these having a high probability that the content is either invalid (i.e., it
was never in IPFS) or the content was no longer available at the time of the
request. As for the cIds in requests, these appear in the full \texttt{url} of
the request that is obtained by concatenating the HTTP host field with the HTTP
target field. Note that the HTTP host in this case can be \texttt{ipfs.io} (i.e., the
gateway host), in which case the cId will appear on the HTTP target; or can be in the
form of \texttt{$<$cId$>$.ipfs.dweb.link}, in which case the cId is in the HTTP
host part. Note as well, these \texttt{url} contain a file system path to the
requested content however, the cId might represent a folder in that path or the
file. In the case a \texttt{url} contains multiple cIds, we only consider the
root cId (i.e., the cId that is in the beginning of the file system path), as
effectively, it is most likely that all cIds found under a root cId are also
provided by the same provider peer. This also groups requests to cIds found
under a root cId to the root cId, which does not alter the frequency of requests
made to a cId, as the root cId always encompasses the remainder cIds.

With this filter, we filter out $41\%$ of all \textsc{GET} operations, where
$17\%$ of these were \textsc{GET} operations that did not succeed and $24\%$ are
requests that did not contain a cId. With this, $47\%$ of the total requests
remained as valid \textsc{GET} operations, which were the ones consider for our
study. Table~\ref{tab:reqs} summarizes the requests we processed for our study.

\begin{table}[t!]
  \centering
  \caption{Requests processed summary.}
  \begin{tabular}{ l | c | r }
    \toprule
    & Number of entries & Percentage\\
    \midrule
    Total & $123,959,912$ & $100\%$  \\
    Out of Format & $2,165$ & $0.001\%$ \\
    Not GETs & $24,298,396$ & $19.602\%$ \\
    All GETs & $82,439,744$ & $80.396\%$\\
    Valid GETs & $58,869,788$ & $47.491\%$\\
    \bottomrule
  \end{tabular}
  \label{tab:reqs}
\end{table}

\subsection{Locating the content providers}

The second step in studying and characterizing the IPFS workload is to gather
information on the providers of the requested content, as to understand where
and by how many peers is the content served. To achieve this, we developed
a simple libp2p program that connects to IPFS and requests all providers of a
given cId. Our program leverages the fact that IPFS uses the same networking software
stack and DHT provided by libp2p, which by default connects to the IPFS network,
to perform \textsc{FindProviders} API calls to libp2p's DHT to fetch
the providers of cIds.

Out of the $58,869,788$ valid \textsc{GET} operations, a total of $4,009,575$
different cIds were requested. We requested the providers of all these cIds
through our libp2p program. We found we were unable to locate the providers of
$54\%$ of all cIds. This can be due to the fact that the content was no longer
available on the network (note that this study was perform some months after the
requests were recorded by the gateway). From the cIds with providers we
discovered $55,830$ different providers however, $59\%$ of these did not have
any addressing information associated to them. This means that the peers storing the  provider
record did not receive any update on the provider (from any kind of traffic)
over a $30$ minute window, as such, the peers storing the provider record
assumed the provider might have moved and has a different address, thus
forgetting the provider's multiaddress. To fetch the multiaddress in these
cases, we queried again the DHT for the multiaddress of the provider, and
managed to find the multiaddress of $4,024$ more providers ($7\%$ more providers
we found initially). Table~\ref{tab:provs} summarizes the numbers of processed
cIds and found providers.

\begin{table}[t!]
  \centering
  \caption{Providers processed summary.}
  \begin{tabular}{ l | c | r }
    \toprule
    & Number of entries & Percentage\\
    \midrule
    Total cIds  & $4,009,575$ & $100\%$  \\
    cIds w/out provider & $2,175,608$ & $54.26\%$ \\
    cIds w/ provider & $1,833,967$ & $45.74\%$ \\
    \midrule
    Providers & $55,830$ & $100\%$\\
    Providers w/out address & $32,968$ & $59\%$\\
    Providers w/ address & $22,862$ & $41\%$\\
    Providers w/ address after find & $26,886$ & $48\%$\\
    \bottomrule
  \end{tabular}
  \label{tab:provs}
\end{table}

\subsection{Analyzing the data}

The final step to study and characterize the IPFS workload is to join both
request and providers data, to map from where in the World are requests being
performed and where in the World is the content provided/available. This
required us to extract geo-locality data from the gathered data. To this end, we
use the MaxMind GeoLite2 Free Geolocation Database\,\cite{maxmind}, that
provides the geo-location of a vast collection of public IP addresses. However,
this database is not complete and may have IP addresses whose geo-location is
unknown. Fortunately, for the request data all IP addresses had geo-locality
information. On the other hand, only $88\%$ of providers with addresses had
geo-locality information.

Note that a provider is identified by a peerId and has multiple
multiaddresses. To get the geo-location information of a provided we had to
extract the (public) IP address of multiaddresses. For multiaddresses that
contained protocols \texttt{ip4} and \texttt{ip6} this procedure is
straightforward. This amounts to $98\%$ all observed multiaddress (excluding
local address, such as \texttt{127.0.0.1} and \texttt{::1}); $0.6\%$ of
multiaddresses were DNS addresses, that we resolved with local DNS resolvers;
and the remainder $1.4\%$ of multiaddress were relay multiaddresses, and hence
the provider did not have a public reachable IP address, which we ignored in this
study. For providers that showed to have multiple locations (probably due to VPN
services), we considered the last observed location. These were just a few cases
that do not affect our study.

We have inserted both datasets into a PostgreSQL database for ease of analysis.
This database has $2$ tables, one containing the requests and another containing
the providers. The requests table stores a request identifier (reqId), the
timestamp the request was originally made to the gateway, the cId requested, and
the location information of the requester. The requests table has as key the
reqId, that is an hash of the request log entry, to avoid processing duplicate
request entries from the log. The providers table stores: the cId provided, the
peerId of the provider, and the location information of the provider. The
provider table has as key the cId and peerId. This uniquely identifies each
provider entry, since each cId can have multiple providers, and a provider can
provide multiple cIds. Note as well that the cId in the providers table is
a foreign key of the cId in the requests table.

By performing a join over the requests and providers table we can compute a mapping from
where requests are performed to where they are provided. Before
presented the results in the next section, we follow by providing some implementation
details to process and find the providers data.

\subsection{Implementation details}  The code and scripts that were used to
process the data for this study can be found in
\url{https://github.com/pedroAkos/IPFS-location-requested-content}. The
processing of data required fine tuning of the parallelization of queries to
IPFS to become timely. This was required because IPFS can take some time to
retrieve the provider records from the DHT; from our study the average latency
was about $6$ seconds, with the maximum latency reaching up to $1.5$ hours; we
made parallel queries to IPFS to fetch provider records. However, libp2p can be
extremely taxing on the network, as a libp2p node can maintain hundreds of
connections and perform thousands of requests. To put this in perspective, a
process executing $100$ queries in parallel to find providers would produce
almost $10,000$ packets per minute. The process to resolve all $4,009,575$
distinct cIds took around $40$ hours.

\section{Results}\label{sec:res} 

In this section we analyze the results from our study to characterize the IPFS workload.
In particular we are interested in answering the following questions:
\begin{enumerate}
    \item \emph{How many requests are made to IPFS on average per day?}
    \item \emph{How is the request frequency distributed over different cIds in the system?}
    \item \emph{How are providers geo-distributed in the system?}
    \item \emph{How is provided content distributed across providers in the system?}
    \item \emph{How does the location of requested content correlate with the location of providers for requested content?}
\end{enumerate}


To answer these question, we analyze the data first from the point of view of
requests by analyzing the requests data extracted from the gateway logs (Section.~\ref{sec:res:reqs}).
We then analyze the data from the point of view of providers using the data we extracted
directly from IPFS
(Section~\ref{sec:res:provs}). Finally, we analyze both requests and providers
data to produce a correlation between the location of requests' origins and
content providers (Section~\ref{sec:res:loc}).

\subsection{Requests}\label{sec:res:reqs}

In this section we analyze the results from the perspective of content fetchers. With
this, we aim to answer the first two questions of our analysis.
\emph{How many requests are made to IPFS on average per day?} and
\emph{How is the request frequency distributed over different cIds in the system?} We begin by answering the first question.

\begin{figure}[t!]
    \centering
    \subcaptionbox{All.\label{fig:reqs:all}}{
      \includegraphics[width=1\linewidth]{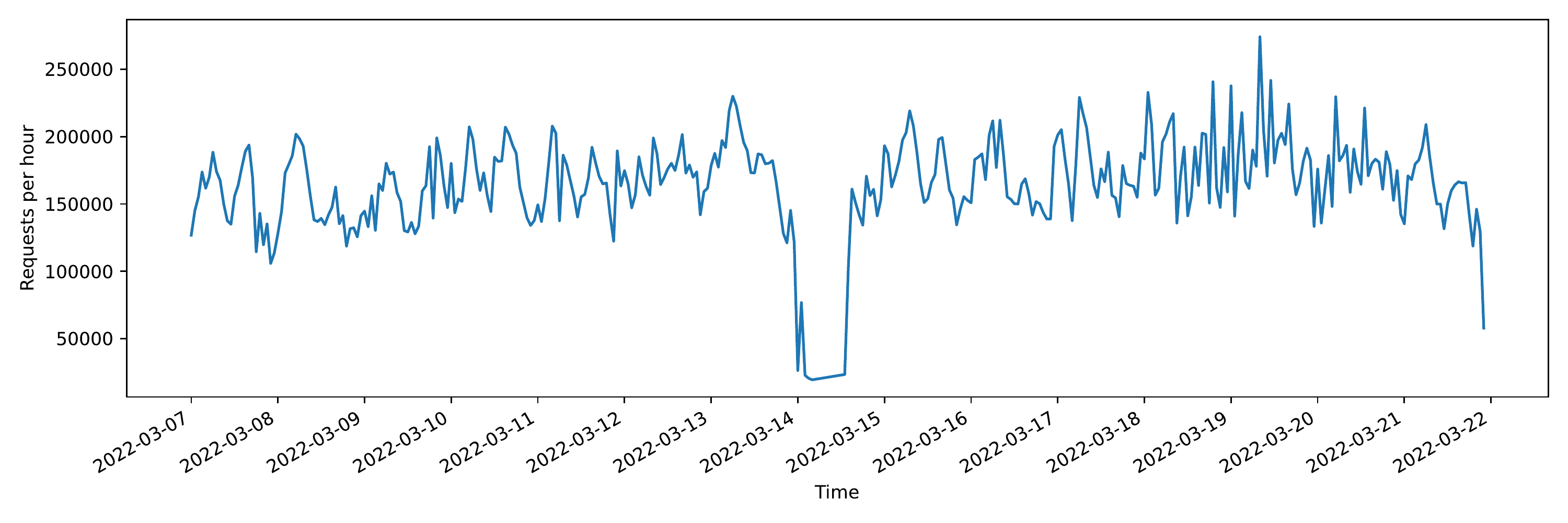}}
  \subcaptionbox{Per continent.\label{fig:reqs:cont}}{
    \includegraphics[width=1\linewidth]{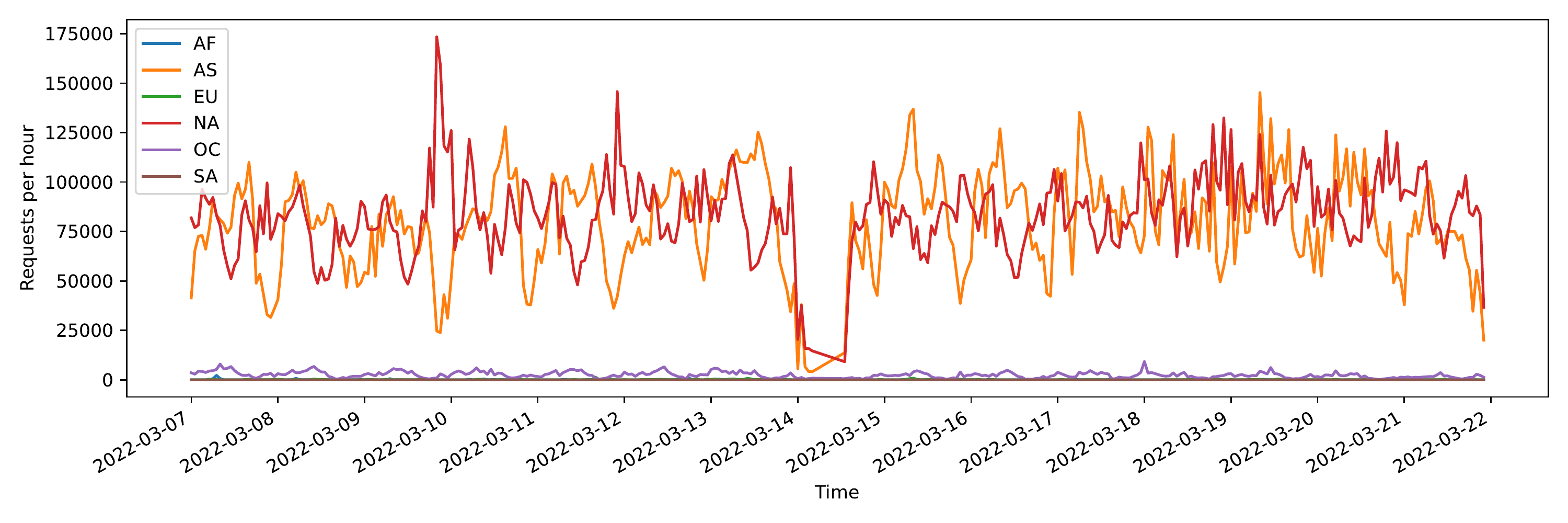}}
  \subcaptionbox{Three days.\label{fig:reqs:cont:day}}{
      \includegraphics[width=1\linewidth]{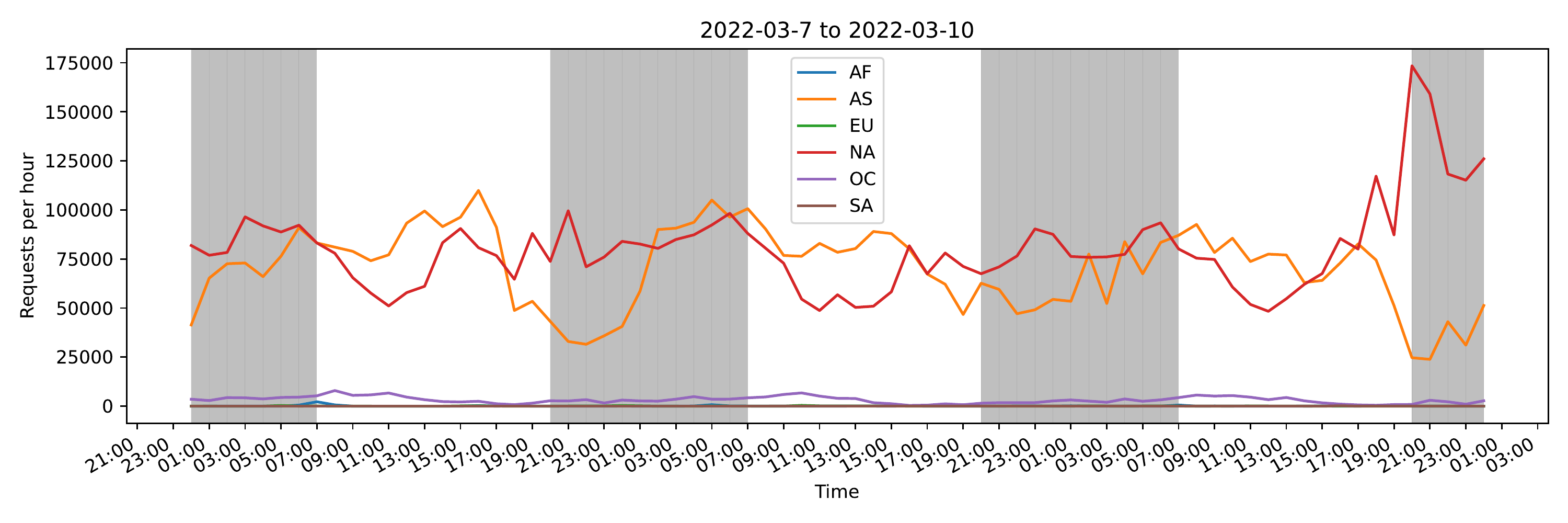}}
  \caption{Requests over time.
  \label{fig:reqs}}
\end{figure}

Figure~\ref{fig:reqs} represents the client requests processed by the gateway per hour.
Notice that Figure~\ref{fig:reqs:all} captures all requests made during the
period of two weeks (x axis), Figure~\ref{fig:reqs:cont} captures the same
requests but characterized by continent, and Figure~\ref{fig:reqs:cont:day}
focuses on the requests characterized by continent for only the first $3$ days of the
analysis period, with the night hours shaded on the gateway's
timezone (GMT-7).

Figure~\ref{fig:reqs:all} shows that, on average, more than $150,000$
requests per hour are made to the IPFS gateway, reaching a maximum of almost
$275,000$ requests per hour. Notice that on day $2022$-$03$-$14$ the
requests suddenly drop. We verified this, and indeed the logs we have from that
day abruptly stop after a few hours (just before the $5$ hour mark). Most likely,
this was due to an  issue with the gateway that day that made it unreachable
for about $9$ hours, which after then resumed processing requests regularly.

From Figure~\ref{fig:reqs:cont} we can see that most of the gateway traffic is
split from North America (NA) and Asia (AS) with more than an average of
$75,000$ requests per hour with origin on each region. The third region with the
most requests per hour is Oceania (OC) with an average of around $2,500$
requests per hour. This is followed by Europe (EU) with an average of around
$85$ requests per hour, Africa (AF) with an average of around $57$ requests per
hour, and South America (SA) with an average of only $3$ requests per hour. From
these results we conclude that this IPFS gateway handles predominantly traffic
from North America and Asia with a high volume of requests. Note that there are
other public gateways located in other regions although, of the most popular
ones, most are located in North America with some located in Europe\footnote{A full list
of public IPFS gateways can be found in
\url{https://ipfs.github.io/public-gateway-checker/}}.

To understand if this high volume of traffic has a day/night pattern, on
Figure~\ref{fig:reqs:cont:day} we plot the requests per day of the first $3$
days of our analysis, and shaded the areas of the plot that represent the night
cycle. Note that the time on the x axis, represents the time at the gateway (on
the North American GMT-7 timezone). From these results, it is not obvious that there
exists a day/night pattern for the North American traffic. However, there is a
slight tendency towards having more traffic during the night, although
marginal. On the other hand, we also noticed that if we shifted the timezone,
the Asian traffic would also not follow the day/night pattern clearly. What is
clear, is that the Asian and North American traffic exhibit somewhat symmetric patterns
over time, whose variations
cancel each other maintaining a high and almost constant load of traffic on the
gateway at all times. This effectively shows that the IPFS network is never idle.


\begin{figure}[t!]
    \centering
    \subcaptionbox{ECDF.\label{fig:cids:all:ecdf}}{
      \includegraphics[width=0.45\linewidth]{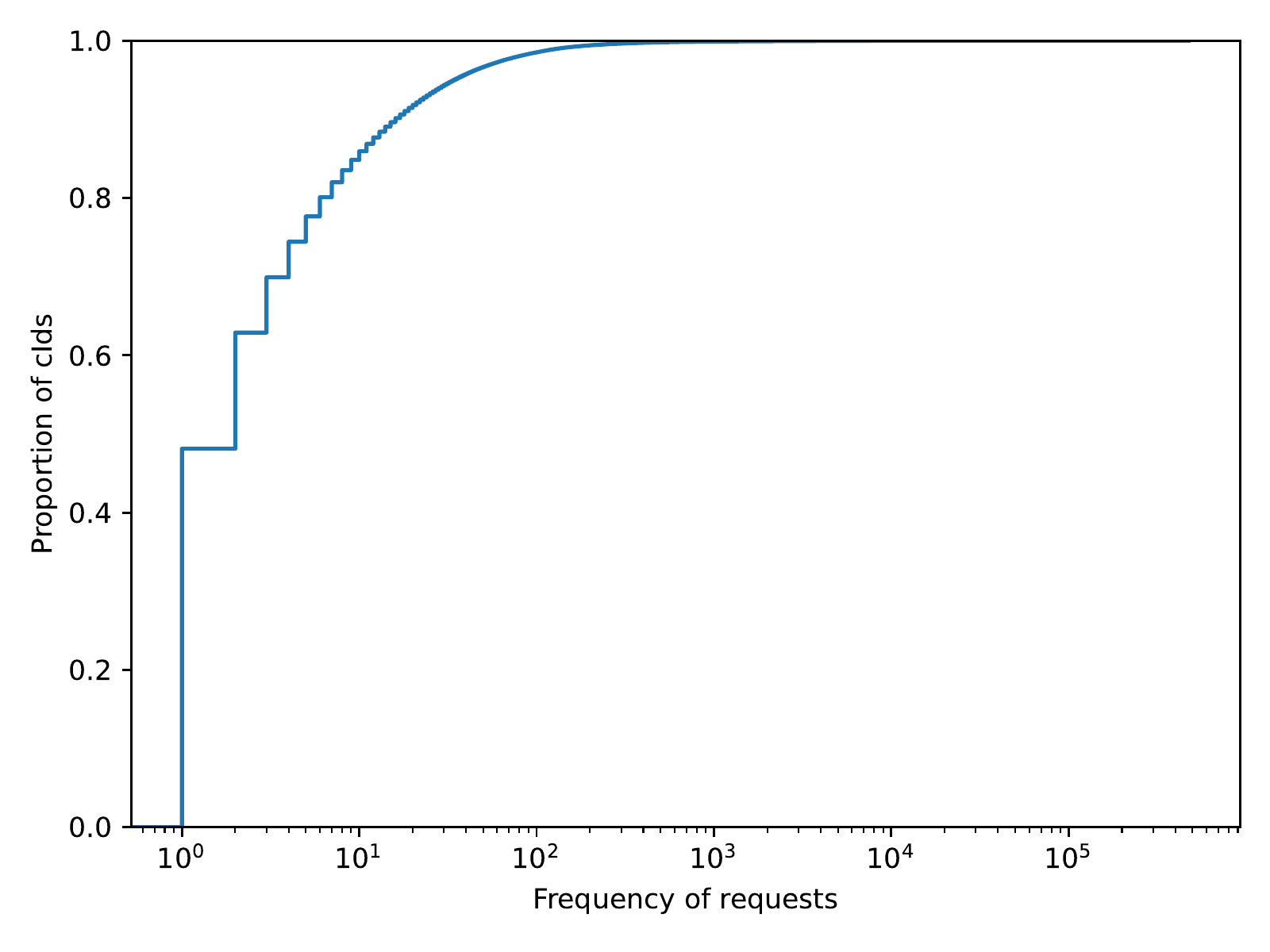}}
  \subcaptionbox{Distribution.\label{fig:cids:all:scatter}}{
    \includegraphics[width=0.45\linewidth]{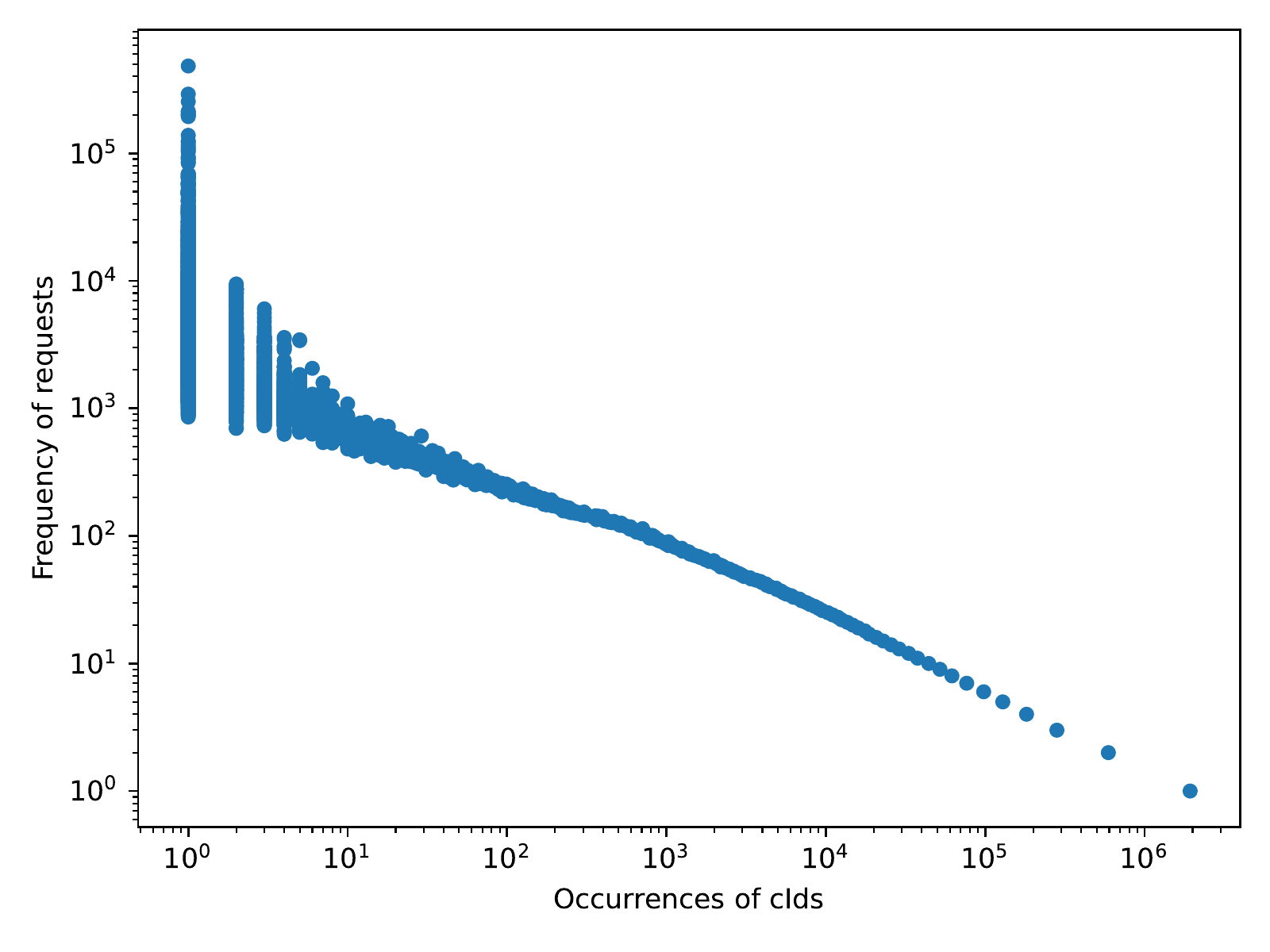}}
  \caption{All requested cIds frequency.
  \label{fig:cids:all}}
\end{figure}

\begin{figure}[t!]
    \centering
    \subcaptionbox{ECDF.\label{fig:cids:cont:ecdf}}{
      \includegraphics[width=0.45\linewidth]{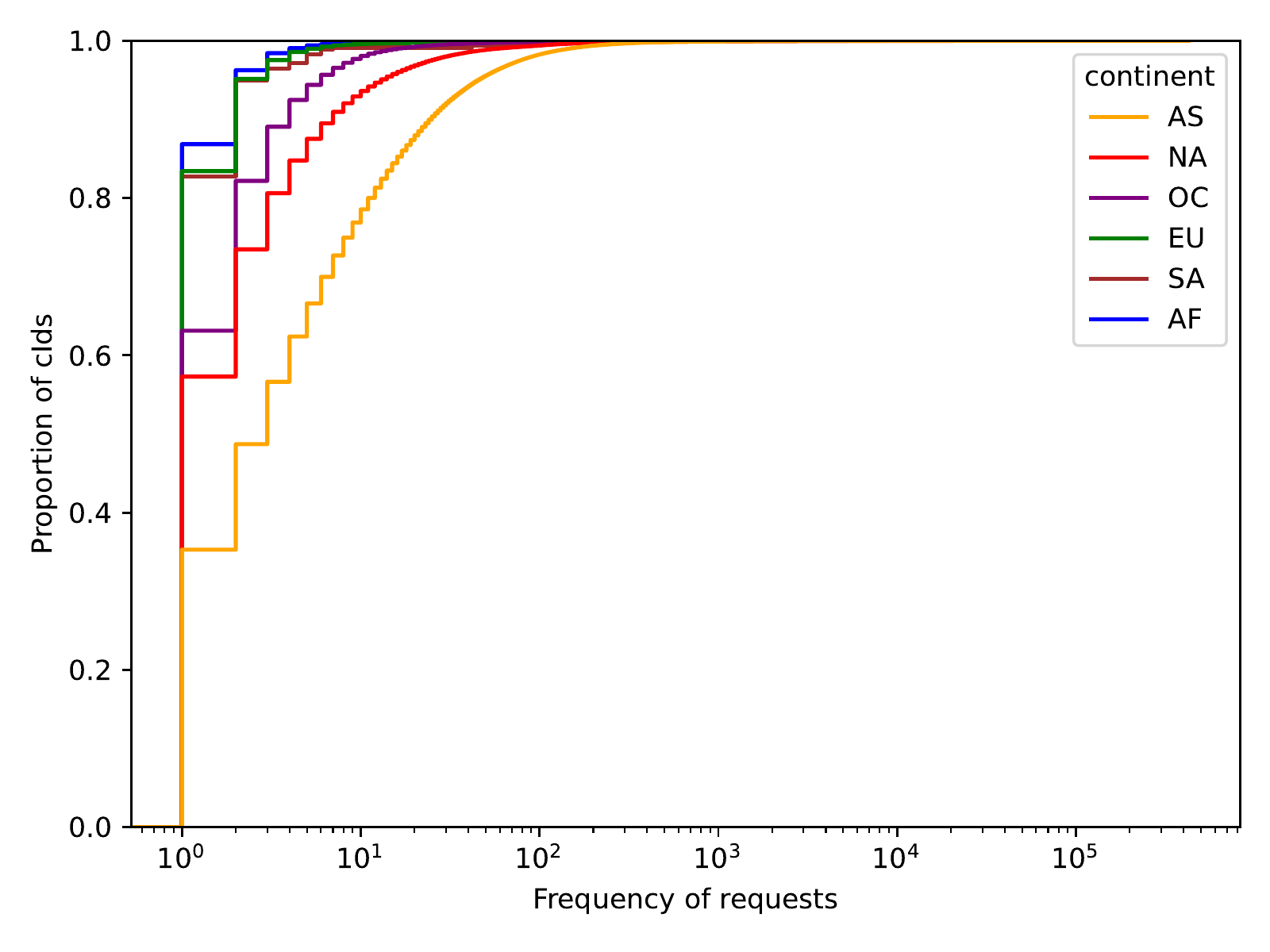}}
  \subcaptionbox{Distribution.\label{fig:cids:cont:scatter}}{
    \includegraphics[width=0.45\linewidth]{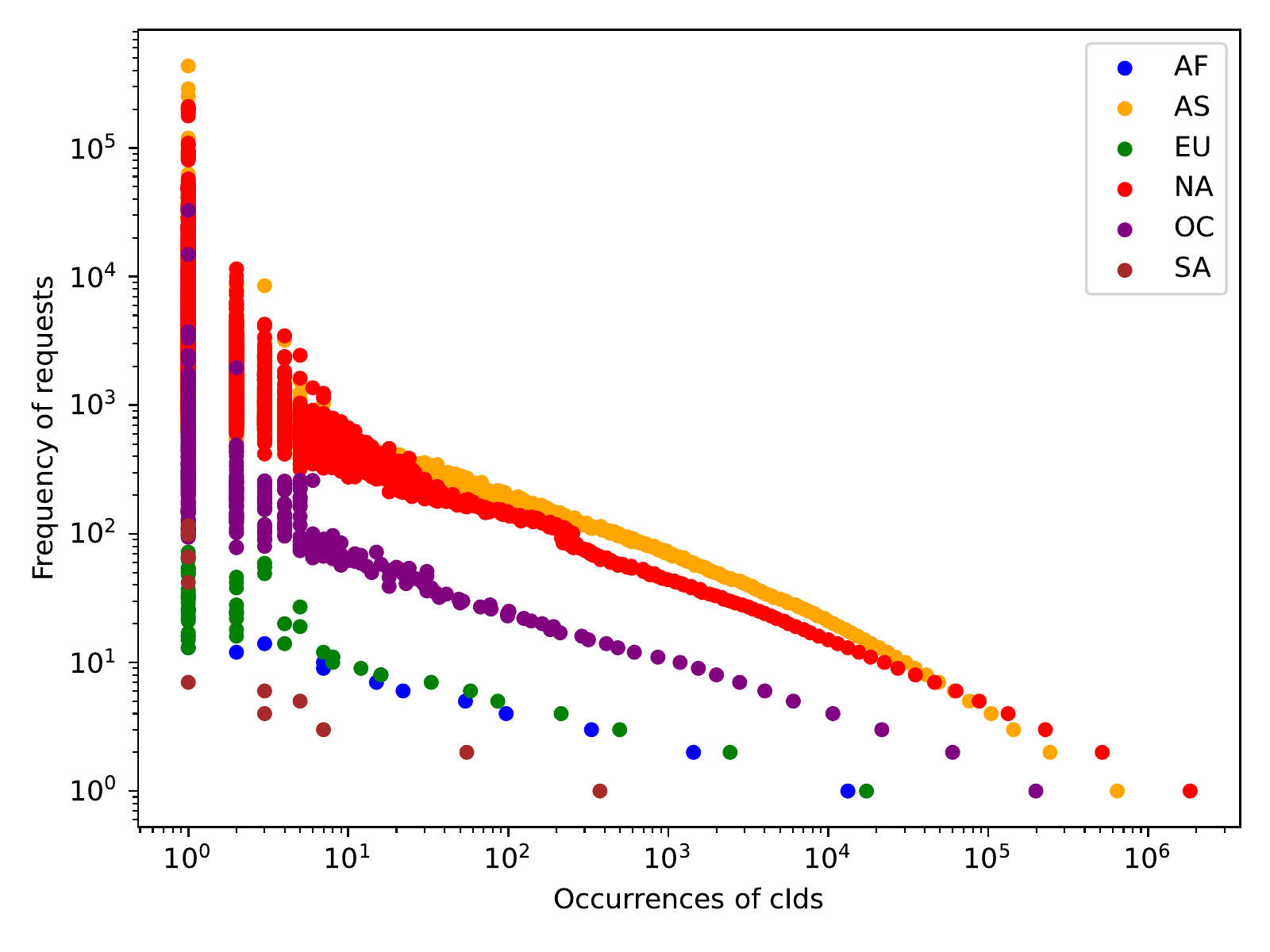}}
  \caption{Per continent requested cIds frequency.
  \label{fig:cids:cont}}
\end{figure}

Figures~\ref{fig:cids:all}~and~\ref{fig:cids:cont} represent the frequency of
requests performed for a cId (i.e., how many times was a cId requested from the
gateway by a user). These results serve to answer question $\#2$: \emph{How is
the request frequency distributed over different cIds in the system?}

Figure~\ref{fig:cids:all:ecdf} shows an ECDF for all requested cIds. Notice that
the x axis (representing the frequency of requests) is in logarithmic scale. The
y axis captures the proportion of requested cIds with at most that amount of
accesses. We notice that almost half of all cIds are only requested once. After
that, the frequency increases with decreasing increments on the proportion of
cIds, where we that about $90\%$ of all cIds are requested at most $10$ times and
about $99\%$  of all cIds are requested at most $100$ times.
Figure~\ref{fig:cids:all:scatter} complements the ECDF showing the distributions
of frequency of requests (shown on the y axis) over the number of cIds (shown on
the x axis). Each point in this distribution represents how many cIds where
requested how many times. Note that both axis of this figure are in logarithmic
scale. From this figure we can see the tendency on the frequency of requests
over the number of cIds, which resembles a typical Zipf distribution.
Table~\ref{tab:top10} summarizes the frequency of the top $10$ requested cIds.
We can further add, that most of these top $10$ most requested cIds were
Non-Fungible Tokens (NFT) related data, suggesting that this is a primary use
case for IPFS.

Figure~\ref{fig:cids:cont} shows the same information but characterized by the
following regions: Africa (AF), Asia (AS), Europe (SA), North America (NA),
Oceania (OC), and South America (SA). Figure~\ref{fig:cids:cont:ecdf} shows an
ECDF for the frequency of requested cIds (on the x axis in logarithmic scale)
over the proportion of requests (on the y axis), discriminated by region. We
notice that almost $60\%$ of requests originating from Asia, request at most the
same cId thrice, whereas $60\%$ of requests originating from North America
request the same cId only once. This shows that content requested from Asia has
a higher popularity (i.e., the same cIds are requested more often) than in the
remainder of regions. Figure~\ref{fig:cids:cont:scatter} complements the ECDF
with the distributions of frequency of requests (shown on the y
axis in logarithmic scale) over the number of cIds (shown on the x axis in
logarithmic). However, it shows that all regions seem to present a similar Zipf
distribution albeit, with different proportions, that is proportional to the
request rate originating from each region.

\begin{table}[t!]
  \centering
  \caption{Top 10 summary of data in descending order.
  The first column represents the amount of requests to each cIds.
  The second column represents the amount of replicas of each cIds.
  The third column represents the amount of different cIds provided by each provider node.
  Each column is independent, encoding different cIds and providers.}
  \begin{tabular}{  c | c | c  }
    \toprule
    Requested cIds & Replicated cIds & cIds per Provider \\
    \midrule
     $482,620$ & $12,610$ & $869,734$   \\
     $290,485$ & $5,274$ & $213,837$   \\
     $254,132$ & $4,663$ & $202,947$   \\
     $213,019$ & $2,047$ & $200,373$   \\
     $209,913$ & $1,876$ & $176,180$   \\
     $203,510$ & $1,822$ & $174,803$   \\
     $199,628$ & $1,404$ & $173,315$   \\
     $198,500$ & $1,372$ & $144,422$   \\
     $193,432$ & $1,283$ & $108,023$   \\
     $138,084$ & $1,272$ & $107,885$  \\
    \bottomrule
  \end{tabular}
  \label{tab:top10}
\end{table}

\subsection{Providers}\label{sec:res:provs}

In this section we analyze the results from the perspective of providers. With
this, we aim to answer questions $\#3$ and $\#4$ of our analysis. \emph{How
are providers geo-distributed in the system?} and \emph{How is provided content distributed across providers in the system?}

The first question is answered by the results presented in
Table~\ref{tab:provs:cont}. which shows the number of providers per continent.
Notice that North America (NA) has almost as many providers as Europe (EU) and
Asia (AS) together. This shows, that North America composes the largest portion
of the content providers on the IPFS system (which is corroborated with previous
studies\,\cite{studyipfs1, studyipfs2}). Furthermore, notice the last two
columns in the table, that represent providers that only had a relay address
(labeled as Rl), meaning they where behind a NAT without a public address; and
providers whose location information was unknown (labeled as Un), meaning there
was no entry on the MaxMind database for those providers public IP address. As
the location of Rl nodes is also unknown, from this point on all Rl nodes are
considered as belonging to the Un category.

\begin{table}[t!]
  \centering
  \small
  \caption{Providers geo-distribution.}
  \begin{tabular}{ l | c | c | c | c | c | c | c | c | c }
    \toprule
    & AF & AS & EU & NA & OC & SA & AN & Rl & Un \\
    \midrule
    Providers & $40$ & $4959$ & $5789$ & $10983$ & $431$ & $104$ & $1$ & $2473$ & $689$ \\
     \bottomrule
  \end{tabular}
  \label{tab:provs:cont}
\end{table}

\begin{figure}[t!]
    \centering
    \subcaptionbox{All.\label{fig:cid_replica:all}}{
      \includegraphics[width=0.45\linewidth]{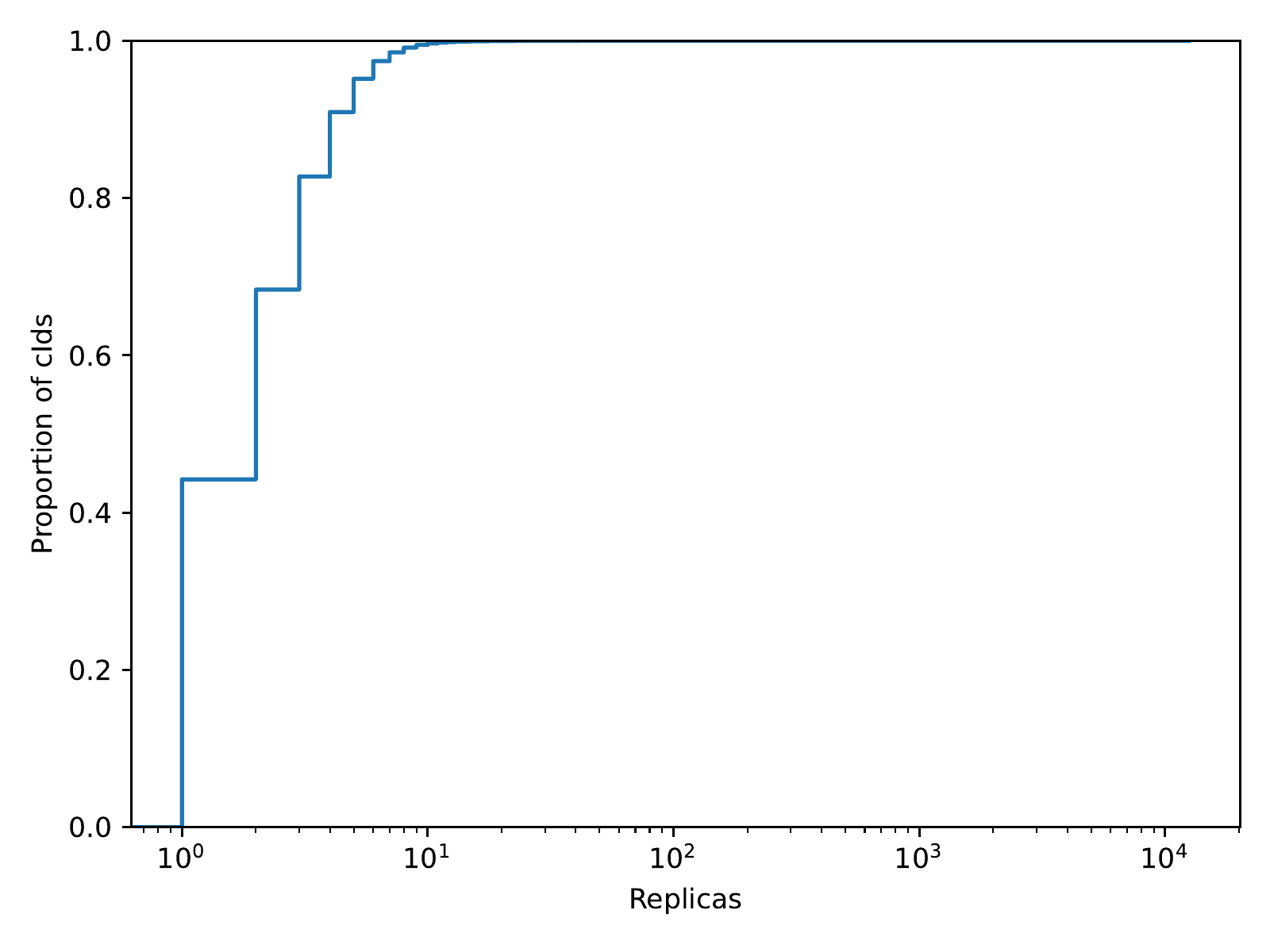}}
  \subcaptionbox{Per continent.\label{fig:cid_replica:cont}}{
    \includegraphics[width=0.45\linewidth]{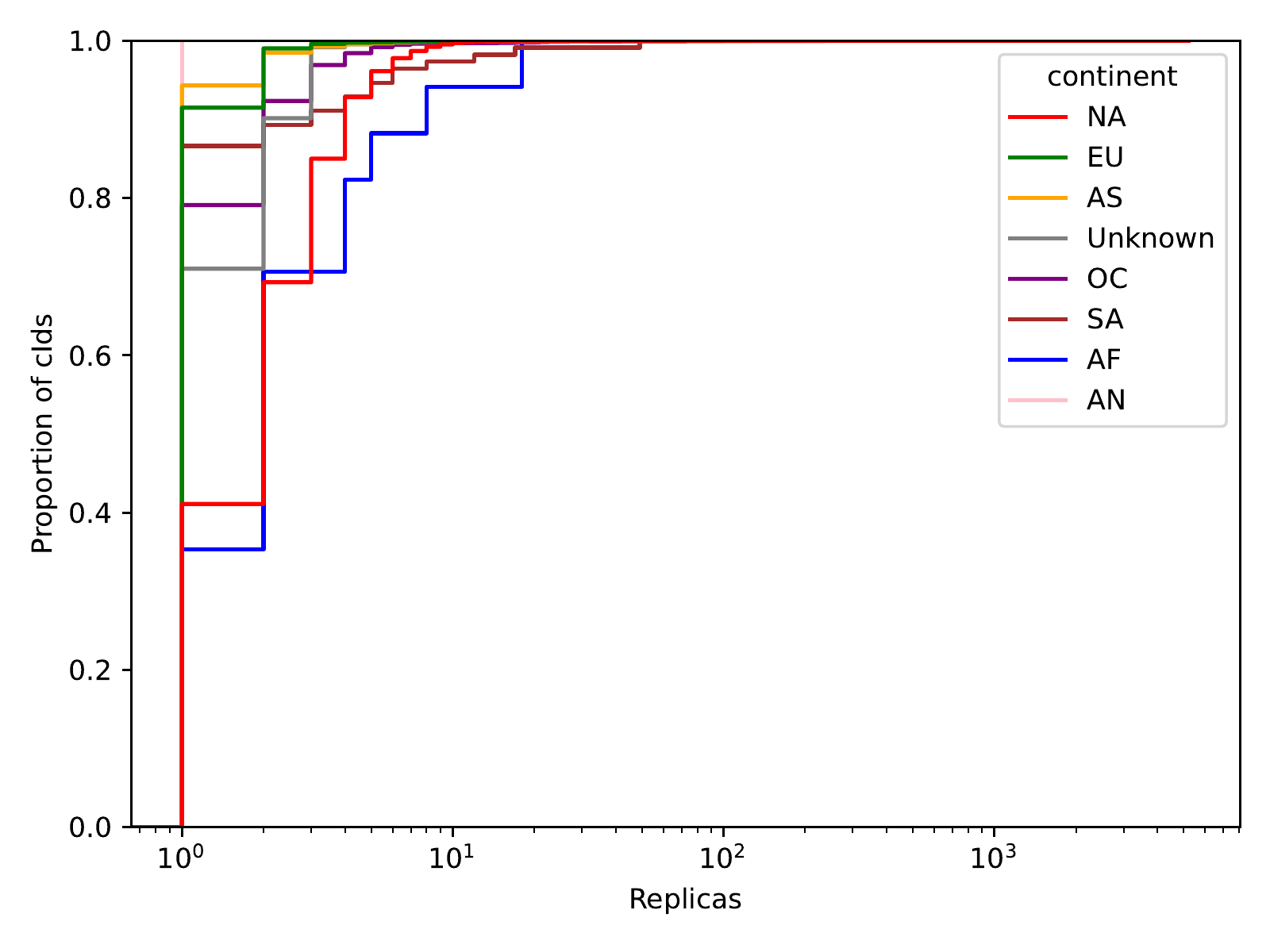}}
  \caption{cId replicas.
  \label{fig:cid_replica}}
\end{figure}

\begin{figure}[t!]
    \centering
    \subcaptionbox{All.\label{fig:prov:all}}{
      \includegraphics[width=0.45\linewidth]{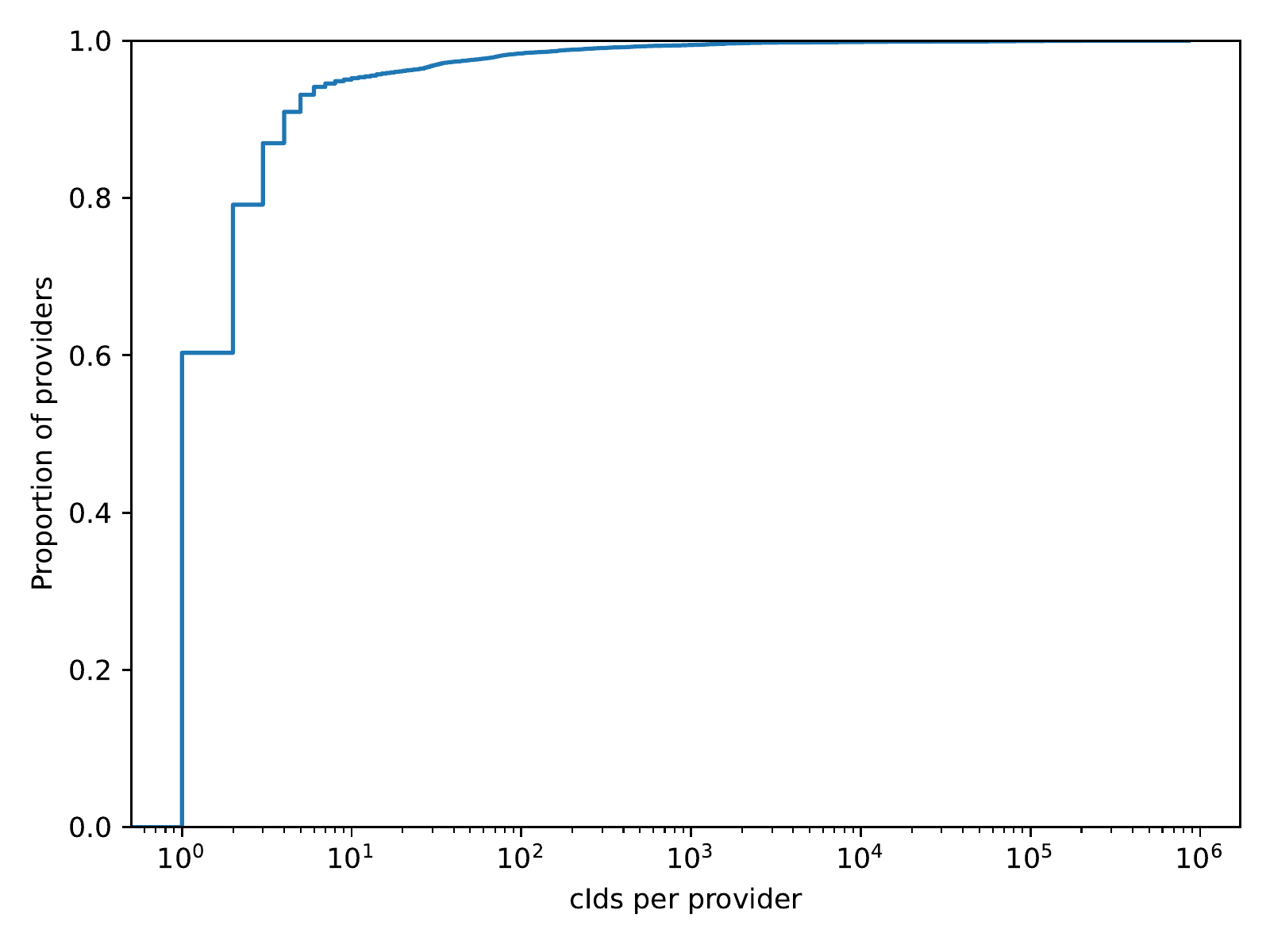}}
  \subcaptionbox{Per continent.\label{fig:prov:cont}}{
    \includegraphics[width=0.45\linewidth]{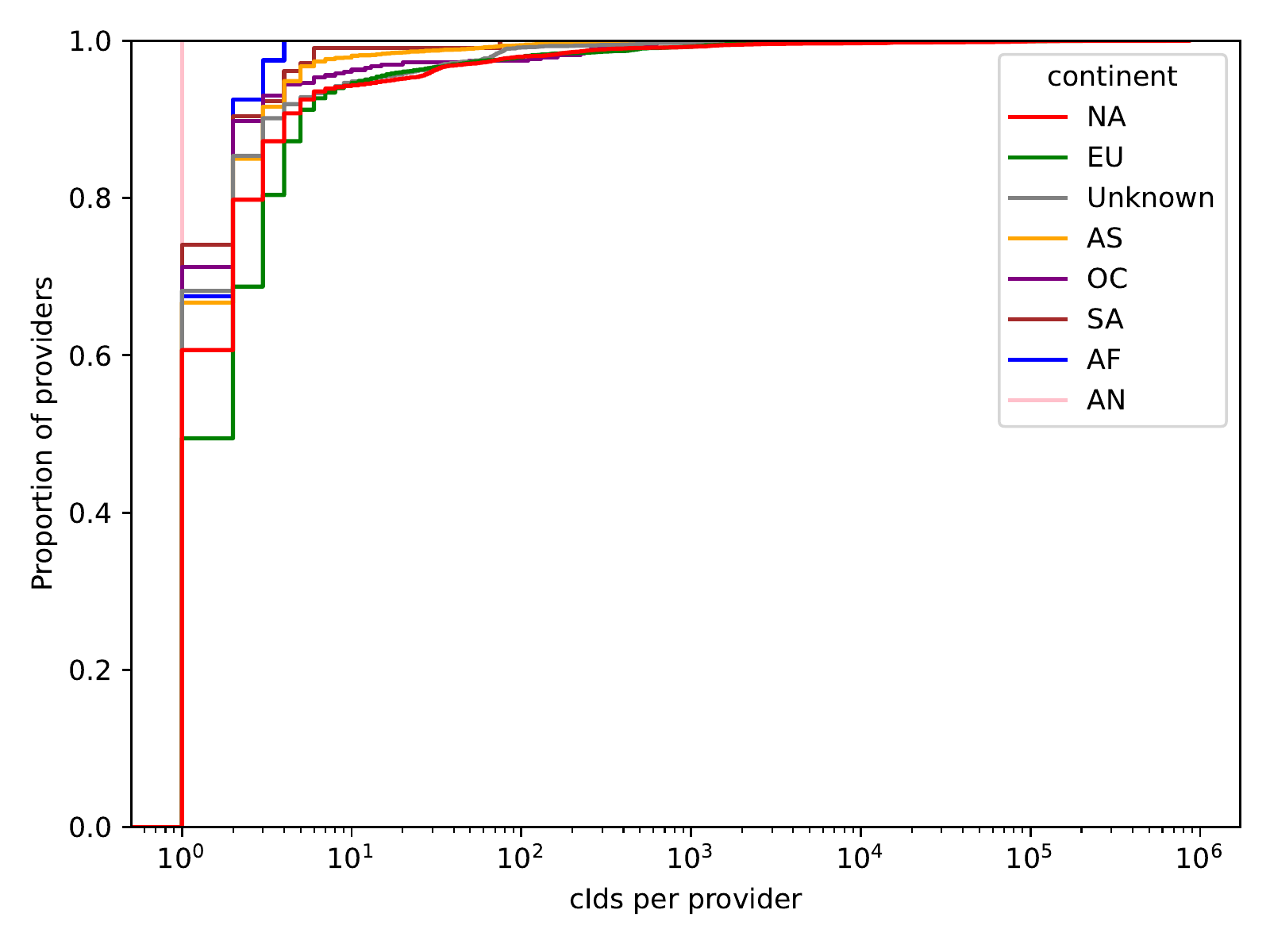}}
  \caption{cIds per provider.
  \label{fig:prov}}
\end{figure}

To answer question $\#4$: \emph{How is provided content distributed across providers in the system?}
we analyze both the amount of content replication
in the system (Figure~\ref{fig:cid_replica}) and the amount of (different) content provided by
each individual provider node in the system (Figure~\ref{fig:prov}).

Figure~\ref{fig:cid_replica} analyzes the amount of replicated content over
different providers. Figure~\ref{fig:cid_replica:all} shows an ECDF that
captures the amount of replicas (on the x axis in logarithmic scale) that were
found for the proportion of cIds in the system (on the y axis). We note that
almost $70\%$ of all cIds in the system are replicated at most twice (i.e.,
provided by at most two different provider nodes in the system). Only a very
small proportion of cIds are replicated by a large number of providers.
Table~\ref{tab:top10} summarizes the amount of replicas of the top $10$
replicated cIds, where, after looking into these cIds, we found that most of
them are IPFS manual pages. We verified if theses highly replicated cIds where
also the most requested cIds and found that this was not the case. In fact, the
top $10$ requested cIds are not very replicated in the system, having only a few
providers (only $3$ of these cIds have more than $10$ providers, with the most
provided being by $29$ and $28$ peers). Figure~\ref{fig:cid_replica:cont} breaks
down the cId replicas by region. Here we notice that Africa has the most
replicas of cIds although, this does not represent a large number as there are
only a few providers in that region. Although it is not visible in the plot,
there is a small percentage of cIds that is highly replicated in North America.
This is not surprising, as North America has the largest number of content providers.
Nevertheless, these results suggest that there is a very limited high
availability of provided content through replication in the system. This can be
mostly explained by the way content is replicated, where there needs to be an
explicit (re)provide action by the user after fetching a copy of the content from
other provider(s).

Figure~\ref{fig:prov} shows the amount of different cIds each provider provides.
Figure~\ref{fig:prov:all} presents an ECDF of the proportion of providers (on
the y axis) that provide different amounts of cIds (on the x axis in logarithmic
scale), here we can see that $60\%$ of providers only provide a single cId. We
also note that less than $10\%$ of providers provide at least $10$ cIds, with a
very small proportion of providers providing at least $1,000$ cIds. The
providers that provide more cIds amount to the largest part of provided cIds,
meaning that most cIds are provided by the same small set of providers. This is not
surprising, as there are pinning services for IPFS content (services that hosts
content for users on IPFS in servers controlled by the pinning service provider
for a fee). Table~\ref{tab:top10} summarizes the top $10$ providers with the
most cIds. Some of these providers had DNS multiaddress, which we verified
pointed to DNS records suggesting these providers belonged to
\texttt{nft.storage}, which is a popular storage service for NFT content in
IPFS. Figure~\ref{fig:prov:cont} analyzes the proportion of providers (on the y
axis) that provide different amounts of cIds (on the x axis in logarithmic
scale) categorized by continent, which shows that the large providers are mostly
located in North America (NA), Europe (EU), and Oceania (OC). The fact that the
biggest portion of cIds is provided only by a small set of providers suggests
that although IPFS is a decentralized content network system, the content stored
in IPFS presents a hight degree of centralization in this small set of provider
peers.

\subsection{Requested content vs. provided content}\label{sec:res:loc}
Finally, in this section we combine gathered data from the requests and the
providers to obtain a global view of the workload, and to answer the
last question: \emph{How does the location of requested content correlate with the location of providers for requested content?}

\begin{figure}[t!]
    \centering
    \includegraphics[width=0.6\linewidth]{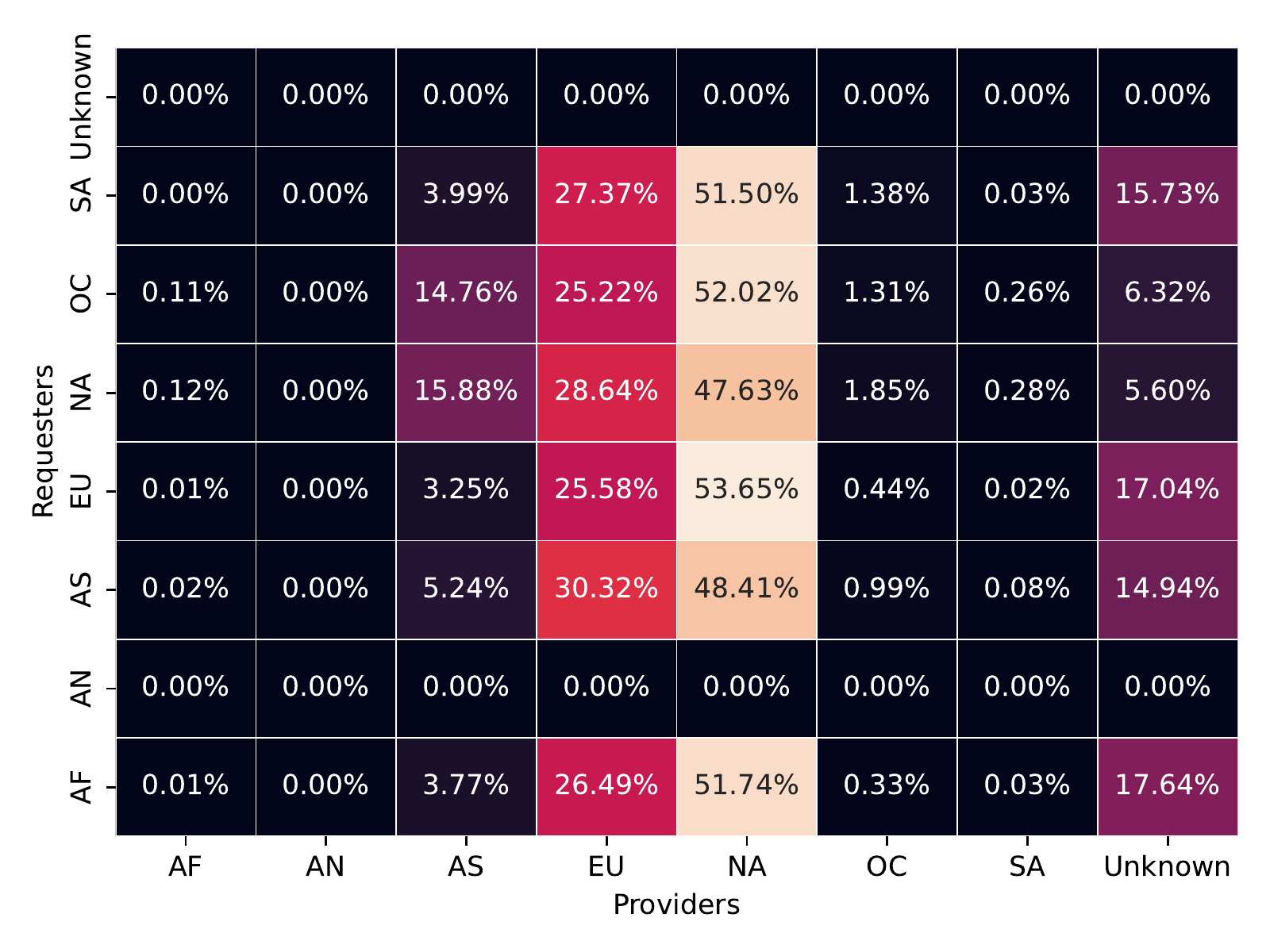}
  \caption{Request locality.
  \label{fig:heatmap}}
\end{figure}

To this end, we matched the request's origin location to the providers location,
generating a heatmap (presented in Figure~\ref{fig:heatmap}) that matches the location from where each
request was made to the location of provider(s) that had the requested content. On Figure~\ref{fig:heatmap} the rows
represent the location of the origin of requests while the columns present the location of
providers. Note that a single request can be served by
multiple locations, as per the cId replication factor we discussed previously.
We normalized the requests per region to eliminate the disparity in quantity of
requests, showing on the heatmap the percentage of all requests made from one
region to any other region (including itself).

By analyzing the heatmap, we notice that vast majority of all requests from all
regions are either provided by providers in North America or in Europe. This is
not surprising as the vast majority of content is located in these  regions.
Furthermore, from this heatmap we can conclude that the North American region
contains the most popular cIds, which from the previous  observation that
requests follow a Zipf distribution, we may also conclude that the Zipf is not
independent per region. Finally, one important observation that can be extracted
from this, is that IPFS is indeed a planetary-scale distributed service, used by
users to share content that is not mostly local, otherwise we should observe
a large majority of content being accessed by users to be served by providers in
the same region (i.e., the heat would be concentrated mostly on the
diagonal of the heatmap)

\section{Related Work}\label{sec:related} 

Measuring and understanding the behavior of large scale and decentralized systems has always been
an important task, with a vast amount of studies being made for
peer-to-peer systems in the early $2000$'s. In particular we highlight two
studies over peer-to-peer systems, one that also analyzes the traffic of large
scale systems, and a second that characterizes the workload of BitTorrent. The
first\,\cite{studyp2p1} analyzes the peer-to-peer traffic across large networks
such as FastTrack, Gnutella, and DirectConnect, from the point of view of a
single ISP. Their findings are similar to ours in the sense that they also
observe most of the traffic being generated by a very small amount of hosts. The
second study\,\cite{studyp2p2} analyzes the popular BitTorrent peer-to-peer
file-sharing system. In this study, the authors perform active measurements in
the systems to obtain data related to upload and download speeds. On the other
hand, our study focuses on other aspects of the system such as the distribution
and popularity of content being published and accessed through the system and
access patterns to that content.

More recent studies on peer-to-peer systems include blockchain systems such as
Bitcoin and Ethereum. Here we highlight two
studies\,\cite{studyblockchain1,studyblockchain2}, that focus on the transactions
made within Bitcoin and Ethereum blockchains. These studies characterize the
performance of blockchain systems but fail to provide insights over system network
properties, such as the number of peers per region or workload distribution over
peers. Our study complements these by focussing on IPFS, an increasingly relevant
solution of the Web3 ecosystem.

Finally, we highlight two more studies\,\cite{studyp2p1,studyp2p2} over IPFS
that are complementary to our own study. The first study\,\cite{studyp2p1} aims
at mapping IPFS, to understand the distribution of peers over regions. Our own
study confirms the findings from this work and complements it with a
characterization on the popularity of content made available by providers. The
second study\,\cite{studyp2p2} focuses on the characterization of Bitswap
traffic inside the system and how does it affect privacy. Our study focuses on
traffic generated outside the system through the main public IPFS gateway.
However, complementing our study with observations extracted from Bitswap
traffic is a goal for future work.

\section{Conclusion}\label{sec:con} 

In this paper we presented a study over the traffic processed by one of the most
popular IPFS gateways to help guide the design of optimizations for future Web3
systems. The traffic showed the amount of requests to IPFS content from various
locations in the World, although mainly from North America and Asia. We
collected the requests and performed searches on the IPFS system to find the
providers, and their location, for each requested content. We analyzed both datasets and found requests follow
a Zipf distribution, where the most popularly requested content is only provided
by a few providers. This means that IPFS is not well balanced, which may lead
to these popular provider nodes to become overly saturated with requests.
From our observations, we propose a research agenda to improve IPFS with the following points:
\begin{itemize}
    \item IPFS needs to remove the load on highly popular providers. To achieve
    this, new incentive mechanism are required to incentivize peers that fetch
    content to also replicate the content.

    \item Because IPFS publishes provider records on the DHT, the load on the DHT
    will not be balanced by an increased number of providers replicating content.
    This is because the records will still be stored in the same peers of the DHT,
    which suggests that a DHT presents a sub-optimal design for load balancing IPFS workload.

    \item Monitoring large scale systems is an important task that needs to be
    done continuously, as user and application patterns might shift. It is essential
    to complement these types of studies with more observations from systems operating
    in the wild. However, to achieve this, robust telemetry protocols are
    required to be integrated in these Web3 systems.
\end{itemize}

\bibliographystyle{plain}
\bibliography{bib}

\end{document}